# ON INTERDEPENDENCE OF THE PALEOMAGNETIC FIELD CHARACTERISTICS

A. Yu. Kurazhkovskii[1], N. A. Kurazhkovskaya[2], and B. I. Klain[3]

*Borok Geophysical Observatory, Institute of the Physics of the Earth, Russian Academy of Sciences, Borok, Yaroslavl Region, 152742 Russia*

[1]E-mail:*ksasha@borok.adm.yar.ru*
[2]E-mail:*knady@borok.adm.yar.ru*
[3]E-mail:*klain@borok.adm.yar.ru*

**Abstract.** The behaviour of the geomagnetic field characteristics (mean values of the paleointensity, amplitude of its variations and reversal frequency) during the last 170 million years was analyzed. It was found that the mean values of the paleointensity were in direct relation to amplitude of its variations and in reverse relation to reversal frequency. The most considerable changes of the Earth's magnetic field characteristics occurred in Early Cretaceous, Middle Paleogene and Neogene. The analysis of conformity of the reconstructed characteristics behaviour of a geomagnetic field to an $\alpha\omega$- dynamo model was made.

## 1. Introduction

The behaviour of the paleomagnetic field is characterized by its value (paleointensity), a direction, and also by periodicity and amplitude of their changes. The most probable source of the Earth's magnetic field is a geodynamo. Numerical modelling of a geodynamo evolution has shown that the geomagnetic field characteristics should change concordantly [*Glatzmaier and Roberts*, 1995; *Anufriev*

*et al.*, 1997]. However, for a long time reconstructions of the paleointensity behaviour and reversal frequency did not allow detecting the correlation between these parameters [*Petrova*, 1989; *Pecherskii*, 1998].

By the present various characteristics of the geomagnetic field have been reconstructed to various details and precision. Thus, the scale of geomagnetic polarity is mainly conventional and is confirmed by the data from various sources. There exist simultaneously several alternative conceptions concerning the paleointensity behaviour [*Selkin and Tauxe*, 2000; *Tarduno et al.*, 2001; *Heller et al.*, 2003]. The patterns of temporal changes in the amplitude and periodicity of the paleointensity variations have not been studied yet.

The works by *Tarduno et al.* [2001] and *Kurazhkovskii et al.* [2004] demonstrated that rather high values of paleointensity existed in Cretaceous Normal Polarity Superchrone. More detailed investigations of a Cretaceous geomagnetic field by *Kurazhkovskii et al.* [2007] have shown that high values of paleointensity were occured not only during Normal Polarity Superchrone. In their further investigations *Tarduno et al.* [2006] assumed the existence of an inverse relationship between the geomagnetic field intensity and reversal frequency. The reverse dependence between frequency of inversion and paleointensity was observed in the interval of 22-34 million years. [*Constable et al.,* 1998]. The data on the value of the geomagnetic field presented in the works by *Kurazhkovskii et al.* [2004] and *Tarduno et al.* [2006] were insufficient for the comparison of the behaviour of paleointensity and frequency of inversion during the past 170 million years. Recently we were obtained a series of new continuous fragments of the behaviour of paleointensity of Upper Jurassic - Cretaceous using sedimentary rocks of the Russian plate and adjoining territories. It has allowed expanding considerably a volume of the information on average (for a geological age) values and variations of the geomagnetic field intensity.

In this paper the relationship between characteristics of the geomagnetic field existing in last 170 million years is investigated on the basis of generalization of the world data received on thermomagnetized rocks, and the authors' data on sedimentary rocks.

**2. Analyzed material**

In order to understand the behaviour of the geomagnetic field characteristics the reconstructions of the paleointensity fragments during the Jurassic - Cretaceous period were made using sedimentary rocks and partly published by *Kurazhkovskii et al.* [2007]. Some fragments of the paleointensity behaviour obtained on the basis of sedimentary rocks are published for the first time. The results of the paleointensity determinations using thermomagnetized rocks are taken from the database [*http // wwwbrk.adm.yar.ru/palmag/index.html*] (DB). The frequency of inversions was determined using the scale of geomagnetic field inversions from [*Supplements…, 2000*]. The results of numerical modelling of the dynamo evolution were taken from the papers [*Glatzmaier and Roberts*, 1995; *Anufriev et. al.*, 1997].

The collections of sedimentary rocks sampled from the Russian plate and adjacent territories were used for the reconstruction of the geomagnetic field intensity. Earlier these collections were used to draw a regional magnetostratigraphic scale of Middle Jurassic - Cretaceous [*Guzhikov*, 2004]. For construction of the paleointensity the collections (doubles of samples) were selected from this material satisfying to following requirements:

1. Samples were accumulated in recovery conditions (gray color) of the epicontinental seas. The magnetite was the main carrier of samples magnetization.
2. According to the tests of *Guzhikov et al.* [2003] the magnetization of these rocks was of orientational nature.
3. The dependence of the laboratory remanent orientated magnetization - $I_{rd}$ on a magnetic field value was close to linear one.
4. Magnetomineralogical composition of the samples did not vary during a laboratory

redeposition.

5. The recovery primary remanent magnetization - $I_n$ (termal cleaning) was achieved at the temperature should not exceed 350°C.

The collections satisfying to these requirements were used for the further examinations. The samples underwent thermal cleaning (at temperatures 200°C - 350°C), then their $I_n$ was measured. Then the redeposition was made. Measurements of $I_{rd}$ of the resedimented samples were made after thermal cleaning. After a laboratory redeposition one more grading of collections was carried out. Collections with the samples changing magneto - mineralogically in the course of redeposition were excluded from consideration. The control of changes in a mineralogical composition was carried out with the help of measurements of the magnetic susceptibility and a remanent saturation magnetization - $I_{rs}$ before and after redeposition. To check the effect of the redeposition procedure on changes in $I_{rs}$ doubles of samples were used which were not involved in determination of the redeposition coefficient.

The paleointensity was calculated using formula: $H/H_o=R_{ns}/R_{ds\ mean}$, where $H_o$ is the value of the modern laboratory magnetic field, $R_{ns}=I_n/I_{rs}$, $R_{ds}=I_{rd}/I_{rs}$, $R_{ds\ mean}$ is the average value of the $R_{ds}$ parameter for one sedimentary formation. The values of all magnetic parameters were measured after the thermal cleaning.

Information on the sedimentary formations, the mean (for geological age) values of the paleointensity and amplitude of its variations are listed in Table. Besides for comparison mean values of the paleointensity determined on the basis of thermomagnetized rocks from the database are presented in the same Table.

At least three facts evidence the adequacy of the performed paleomagnetic determinations. In some cases fragments of the paleointensity behaviour were obtained using samples of various even-aged sedimentary formations that allowed us to confirm the results of some reconstructions by external convergence. After construction of fragments of the paleointensity behaviour the correlation between dynamics of the paleomagnetic and petromagnetic parameters was checked. In all cases the correlation coefficient was obtained close to zero that excluded any influence of magneto - mineralogical composition variations on behaviour of the paleomagnetic parameter ($H/H_o$). In most cases the mean (for geological age) values of the paleointensity obtained on sedimentary and thermomagnetized rocks coincided to a high degree of accuracy (Table).

Table

| Age | Place of sampling | h(м) | n | H/Ho (sedim) | S | H/Ho (DB) |
|---|---|---|---|---|---|---|
| Maastrichtian | Pudovkino, Saratov Region | 22 | 22 | 0.6 | 0.3-1.8 | 0.84 |
| Campanian | Tuarkyr cut, West Turkmenia | 18 | 16 | 0.7 | 0.3-1.5 | 1.3 |
| Santonian | Golubinskaya, Volgograd Region | 15 | 42 | 1.0 | 0.4-2 | 0.7 |
| Coniacian | | | | | | 0.64 |
| Turonian | Borehole 200, Saratov Region | 8 | 8 | 0.8 | 0.6-1.2 | 0.97 |
| Cenomanian | Borehole 200, Saratov Region | 34 | 34 | 0.6 | 0.2-1.8 | 0.48 |
| Albian | Borehole 200, Saratov Region | 43 | 43 | 0.8 | 0.4-3 | 0.53 |
| Aptian | Mar'ino, the Crimea<br>Sengiley, Ul'yanovsk Region | 24<br>7 | 24<br>7 | 0.75<br>0.8 | 0.4-3 | 0.75 |
| Barremian | Borehole 204, Saratov Region<br>Kremenki, Ul'yanovsk Region<br>Yatria, the Near-Polar Urals | 34<br>10<br>21 | 34<br>9<br>21 | 0.7<br>0.8<br>0.7 | 0.4-2.7 | 0.43 |
| Hauterivian | Well 204, Saratov Region<br>Verkhorech'e, the Crimea<br>Yatria, the Near-Polar Urals | 17<br>43<br>27 | 17<br>41<br>27 | 0.50<br>0.5<br>0.5 | 0.2-1.5 | 0.64 |
| Valanginian | Verkhorech'e, the Crimea<br>Yatria, the Near-Polar Urals | 40<br>25 | 40<br>25 | 0.6<br>0.7 | 0.2-1.2 | 0.77 |
| Berriasian | Balky, the Crimea | 28 | 23 | 0.4 | 0.2-0.8 | 0.37 |
| Tithonian | Borehole 120, Saratov Region<br>Gorodishche, Ul'yanovsk Region | 18<br>11 | 52<br>22 | 0.3<br>0.28 | 0.1-0.6 | 0.32 |
| Kimmeridgian | Gorodishche, Ul'yanovsk Region | 18 | 52 | 0.3 | 0.1-0.5 | |
| Oxfordian | | | | | | 0.36 |
| Callovian | Cheremukha river, Rybinsk<br>Dybki, Saratov | 2<br>8 | 10<br>3 | 0.5<br>0.36 | 0.3-1.2 | 0.52 |
| Bathonian | Sokursky Trakt cut, Saratov | 12 | 25 | 0.41 | 0.2-1.2 | 0.36 |
| Bajocian | | | | | | 0.95 |

Comments: h is the power of sediment formations, n is the number of studied layers, S is the amplitude of the paleointensity variations in the $H/H_o$ units.

## 3. Results

The most representative fragments of the paleointensity behaviour obtained on the basis of sedimentary rocks are given on Figure 1. As it is seen from Figure 1, mean values and the amplitude of paleointensity variations increased from the Tithon to Barremian. Besides, the shape of the paleointensity variations changed. In the studied fragments of the Tithon - Hauterivian paleointensity its variations happened with close amplitudes. After Hauterivian an alternation of paleointensity variations of the small amplitude with its bursts occurred. The materials presented on Figure 1 are a visual evidence of the relationship between variations of the geomagnetic field intensity and a pattern of its variations. So, in the Upper Jurassic at low mean values of the paleointensity its variations had small amplitudes. During the Berriasian - Hauterivian the amplitude of variations and the mean values of the paleointensity increased non-uniformly. The paleointensity was high in the Barremian – Santonian, and the alternation of the burst and quiet generation of the geomagnetic field was observed.

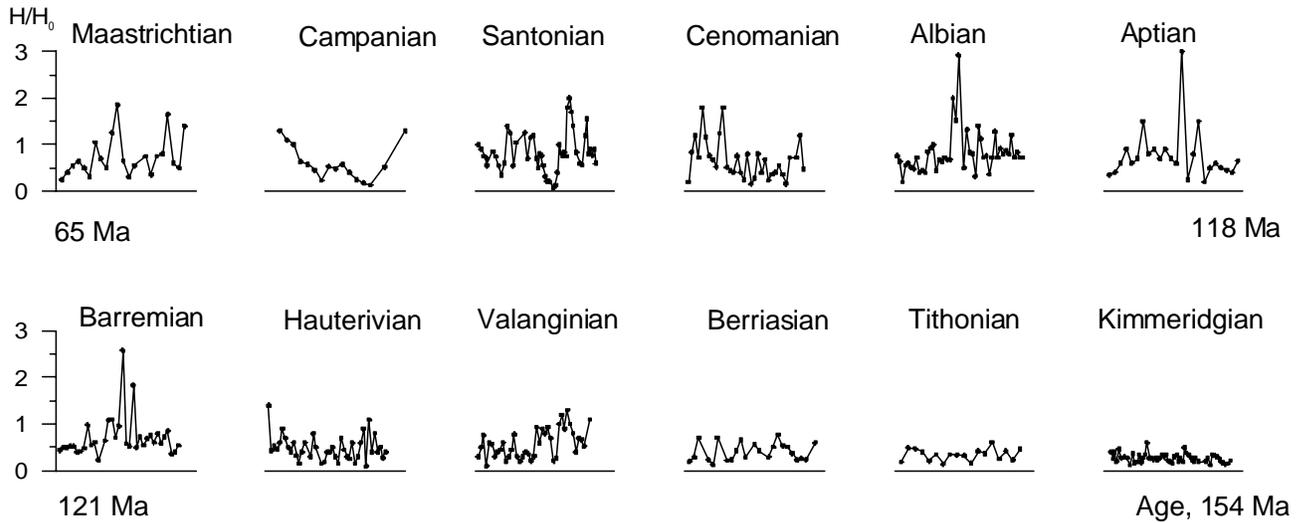

**Figure 1.** The paleointensity behavior fragments in the Jurassic and Cretaceous periods on basis sedimentary rocks.

Figure 2 shows the results of the paleointensity determinations using thermomagnetized rocks. The quantity of these data diminishes non-uniformly from the present to geological past. For the Cenozoic in DB there are from several tens to hundreds of the paleointensity determinations for each geological age. We have assumed that these data can be enough for estimation of the mean values of the geomagnetic field intensity for geological age. But the data from DB are insufficient for estimation of the Jurassic - Cretaceous paleointensity, as for a number of ages they are lacking at all. In this connection a pattern of the paleointensity behaviour during the Jurassic - Cretaceous was made on the basis of average data obtained both on sedimentary and thermomagnetized rocks.

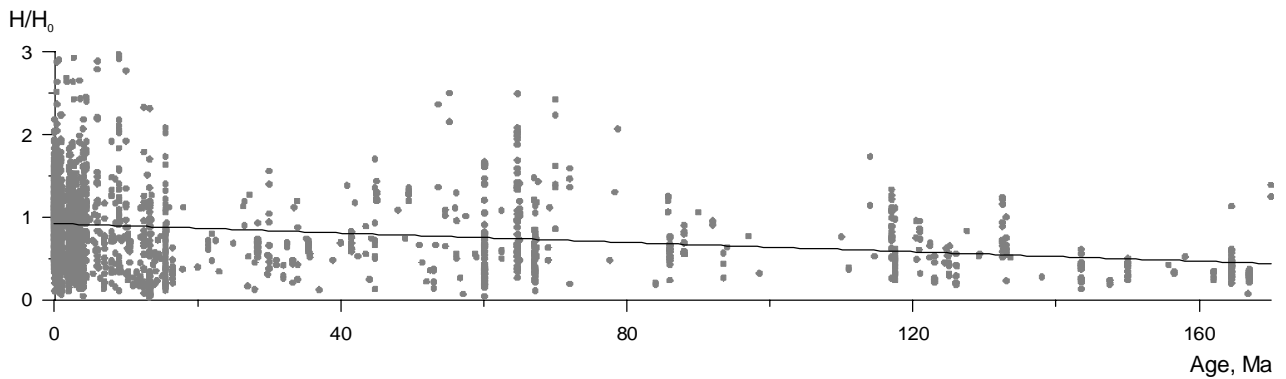

**Figure 2.** Results of the paleointensity determination in the past 170 million years based on the DB. Slope line shows the time trend of paleointensity.

The changes in the paleointensity $H/H_o$ and the reversal frequency $F$ during the last 170 million years are illustrated on Figure 3. Points at the diagram are obtained by averaging the data on paleointensity (dark circles) and the reversal frequency (light circles) of one geological age (stage). Such way of the material representation helps to avoid the confusion of data corresponding to various geological ages, epochs and periods. Intervals of averaging (boundary of geological ages) are cited from [*Supplements…*, 2000]. The values of *F* were obtained by normalization of the number of inversions to duration of a geological age. The information about the number of inversions during a geological age was mainly cited from [*Supplements…*, 2000] also. The Albian – Coniacian interval is an exception. The reversal frequency of the Albian - Coniacian was constructed using the data of the work by *Guzhikov*

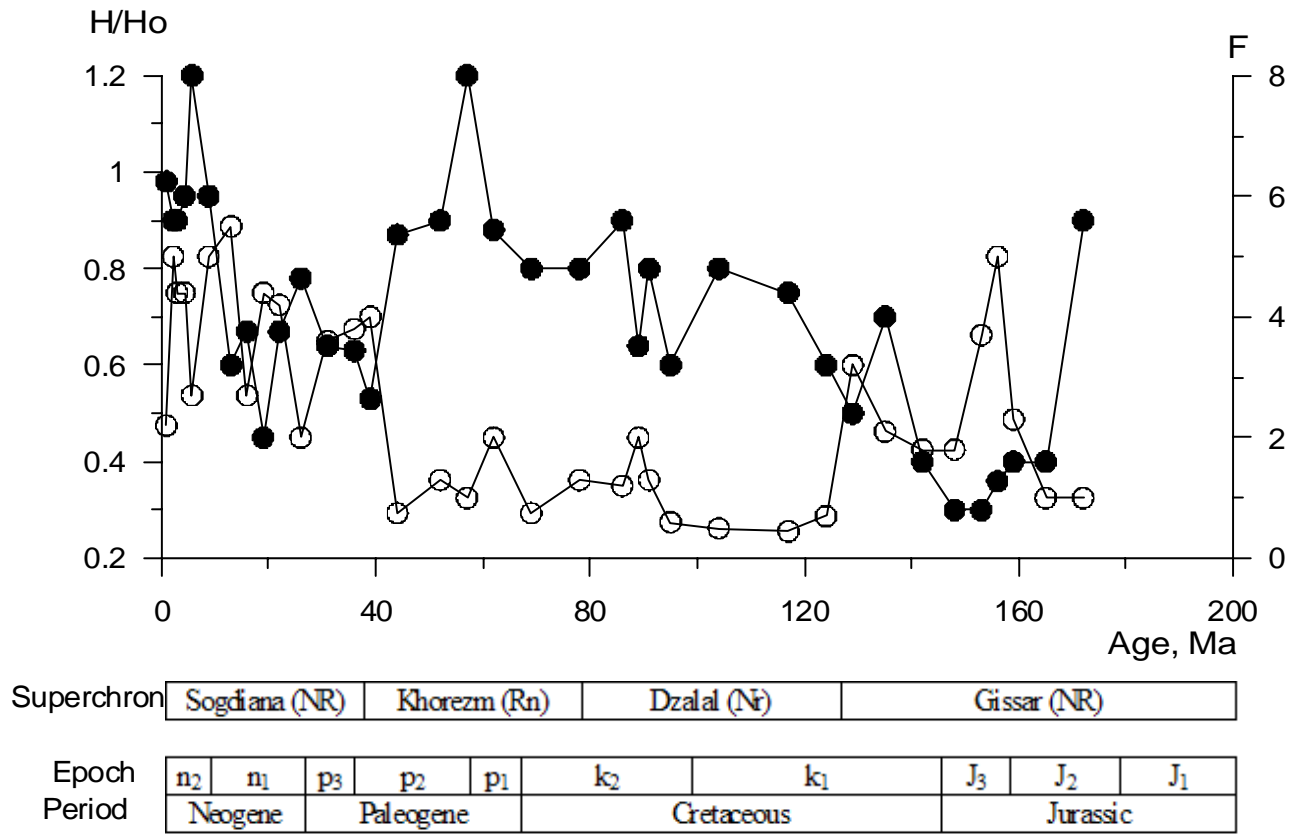

**Figure 3.** Changes of the paleointensity ($H/H_o$) (dark circles) and reversal frequency (*F*) (light circles) during the past 170 million years.

[2004] which are more complete.

As it is seen from Figure 3 the paleointensity behaviour correlates with variations of the reversal frequency of the geomagnetic field. At low reversal frequency rather high values of the geomagnetic field intensity were observed. An increase in the reversal frequency was accompanied by the paleointensity decrease.

An analysis of the behaviour of the geomagnetic field intensity, its variations and reversal frequency testified that the Earth's magnetic field characteristics had changed interdependently. The most considerable changes in behaviour of the geomagnetic field characteristics were marked in the Lower Cretaceous, Middle Paleogene and Neogene (Figure 3). Thus in the Lower Cretaceous the paleointensity increased and the reversal frequency decreased. An inverse process was observed in the Middle Paleogene, the paleointensity decreased, and the reversal frequency increased. During the Neogene the paleointensity increased.

## 4. Comparison of behaviour of the geomagnetic field characteristics with an $\alpha\omega$ - dynamo model

The relationship between the behaviour of the geomagnetic field intensity and the reversal frequency reveals some analogies with the models of the geomagnetic field generation. Thus, in nearly axisymmetric models the magnetic field generation is determined by $\alpha$ - and $\omega$ - effects. The behaviour of the geomagnetic field essentially depends on a dynamo – number ($D$), equal to product of their amplitudes. According to *Sokoloff* [2004] a dynamo - number is proportional to a value of the field intensity. Generation of a sinusoidal field originates at some $D_0$ and at increase of a dynamo - number the fundamental fluctuations frequency increases proportionally to $D^{1/2}$, and further at increase of $D$ at some $D^h$ a regular regime of the field generation is replaced by a chaotic one [*Glatzmaier and Roberts*, 1995; *Anufriev et. al.*, 1997]. It is necessary to mark that since some values of $D_0 < D < D^h$ fluctuations with nonzero value of intensity arise and long-period component of the field appear. At this inverses become asymmetrical and their frequency decreases. Thus it follows from a $\alpha\omega$ - dynamo theory that the regime exists when at increase of the field intensity the reversal frequency declines and the fundamental frequency increases. We think that exactly this regime is realized at generation of the Earth's magnetic field. It is in particular testified by the existence of the paleointensity fluctuations with the period exceeding significantly a fundamental period of the geomagnetic dynamo.

Realized reconstructions do not allow resolving a fundamental frequency of the geomagnetic dynamo, at the same time the dynamics of the other characteristics is available to us. The analysis of the obtained data shows that the increase of the mean values of paleointensity and amplitudes of its variations leads to diminution of reversal frequency of the magnetic field. Thus the behaviour of the reconstructed characteristics of the geomagnetic field is in accordance with an $\alpha\omega$ - dynamo model. An alternation of a quiet geomagnetic field and paleointensity bursts (Figure 1, Barremian - Maastrichtian), and the prevalence of one kind of polarity in Dzalal and Khoresm Superchrones (Figure 2) we are considered as manifestation of the highest possible chaotic character associated with large amplitudes of $\alpha$ - and $\omega$ - effects. It is possible to assume that the further development of the process was limited by intensification of heat outflow into the upper shells of the Earth which began in Early Cretaceous and finished at the end of Middle Paleogene.

## 5. Conclusion

The main conclusions of the work are as follows:

1. Characteristics of the paleomagnetic field (average paleointensity, amplitude of its variations and reversal frequency) changed interdependently.

2. Paleointensity and reversal frequency varied in anti-phase. An increase of the paleointensity mean values was accompanied by an increase of the amplitude of its variations.

3. The most considerable changes in generation of the Earth's magnetic field occurred in Early Cretaceous, Middle Paleogene, and Neogene.

4. The relationship between the geomagnetic field characteristics obtained as a result of analysis of its reconstructions corresponds to an $\alpha\omega$ - dynamo model.

**Acknowledgments.** The authors would like to express their thanks to A.Yu. Guzhikov, M.V. Pimenov and O.B. Yampol'skaya for the collections of sedimentary rock samples and D.D. Sokoloff for a constructive discussion of the manuscript and consultation also.